\documentclass{ifacconf}

\usepackage{graphicx}      
\usepackage{natbib}        
\usepackage{subcaption}
\usepackage{booktabs}
\usepackage{mathrsfs}
\usepackage{bbm}
\usepackage{amsmath}
\usepackage{amssymb}
\usepackage{siunitx}
\usepackage{url}

\usepackage[ruled,vlined,algo2e]{algorithm2e}

\usepackage{longtable}
\usepackage[percent]{overpic}
\newcommand{\capt}[1]{\mdseries{\emph{#1}}}

\usepackage{fancyhdr}
\newcommand{\mytitle}{\textbf{Accepted final version.}
To appear in \textit{Proceedings of the IFAC World Congress, 2026}.\\
\copyright 2026 the authors. This work has been accepted to IFAC for publication under a Creative Commons Licence CC-BY-NC-ND}
\begin{document}
\begin{frontmatter}

\title{Priority-Driven Control and Communication in Decentralized
Multi-Agent Systems\\ via Reinforcement Learning} 


\author[First]{Qingyun Guo} 
\author[First]{Junyi Shi} 
\author[First,Second]{Tomasz Piotr Kucner}
\author[First,Second]{Dominik Baumann}

\address[First]{Department of Electrical Engineering and Automation, Aalto University, Espoo, Finland (e-mail: firstname.lastname@aalto.fi)}
\address[Second]{Finnish Center for Artificial Intelligence, Helsinki, Finland}

\begin{abstract}                
Event-triggered control provides a mechanism for avoiding excessive use of constrained communication bandwidth in networked multi-agent systems. However, most existing methods rely on accurate system models, which may be unavailable in practice. In this work, we propose a model-free, priority-driven reinforcement learning algorithm that learns communication priorities and control policies jointly from data in decentralized multi-agent systems. By learning communication priorities, we circumvent the hybrid action space typical in event-triggered control with binary communication decisions. We evaluate our algorithm on benchmark tasks and demonstrate that it outperforms the baseline method.
\end{abstract}


\end{frontmatter}
\thispagestyle{fancy}
\pagestyle{empty}
\section{Introduction}
Networked Control Systems (NCSs) have attracted considerable attention due to their widespread applications in areas such as smart manufacturing and autonomous driving~\citep{Baumann_2021,Dang2022EventTriggeredMP}. NCSs integrate physical components, including sensors, controllers, and actuators, through a shared communication network. Typically, multiple control loops are established to manage these physical components through the network, making communication a shared and, consequently, limited resource. One strategy to lower communication demand is to adopt event-triggered control (ETC) \citep{Heemels2012AnIT, Miskowicz2015}. In ETC, data are transmitted only when specific events occur, such as significant increases in control error or estimation uncertainty.

Most ETC approaches rely on mathematical system models~\citep{Heemels2012AnIT}, which are often difficult to obtain for high-dimensional and nonlinear systems. A promising alternative is model-free reinforcement learning (RL), which directly learns control and communication policies from data without requiring a system model \citep{FUNK2021100144, SHIBATA2023104307,9830835}. A key challenge in applying RL to learn ETC is handling hybrid action spaces. In this context, each agent must decide at each time step whether to communicate (a discrete action) and determine the appropriate control signal (a continuous action). Due to the complexity of hybrid action spaces, most existing methods do not jointly optimize control and communication policies \citep{8386658,8663430,8713542}, which may lead to sub-optimal results as the separation principle does not necessarily hold for ETC \citep{5990925}.

Beyond this limitation, most RL-based ETC algorithms still focus on single-agent systems \citep{Baumann2018DeepRL, FUNK2021100144, Dang2022EventTriggeredMP}. Few works extend to multi-agent systems, which typically adopt centralized training with global information and assume periodic information exchange during training, even though communication is limited during execution. Under this paradigm, the multi-agent proximal policy option-critic (MAPPOC)~\citep{pmlr-v211-kesper23a} algorithm leverages hierarchical RL to learn continuous control and discrete communication policies jointly. The multi-agent deep deterministic policy gradient (MADDPG) is used to learn ETC with the centralized training and decentralized execution (CTDE) paradigm, which is applied to real robots~\citep{SHIBATA2023104307}. These works assume that during training, there is sufficient bandwidth for all agents to share information periodically. In practice, this assumption is rarely met. Besides, by making discrete communication decisions, in the worst case, also during execution, all agents may communicate at once and overload the network \citep{mager2022scaling}.

\begin{figure}[tbp]
\centering
    \includegraphics[width=0.8\linewidth]{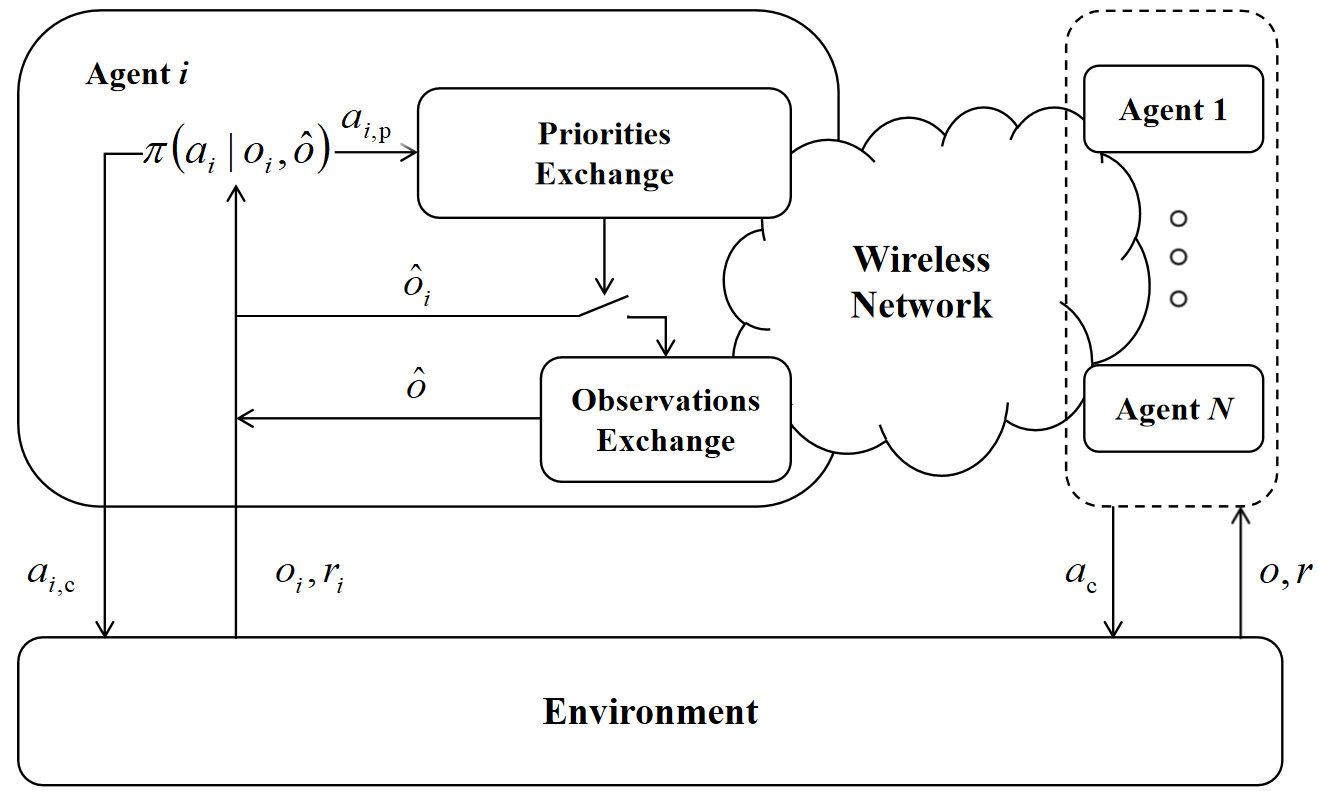}
    \caption{Our framework introduces priority-based control and communication in a decentralized multi-agent setting. \capt{Each agent \(i\) uses its local observation \(o_i\) and received information \(\hat{o}\) to learn the policy \(\pi(a_i \mid o_i,\hat{o})\). Agent \(i\) then transmits its communication priority \(a_{i,\mathrm{p}}\) through the wireless network. The \(L\) highest-priority agents are allocated bandwidth to broadcast their observations \(\hat{o}\). Each agent applies \(a_{i,\mathrm{c}}\) and receives reward \(r_i\) and a new local observation \(o_i\).}}
    \label{fig:sys}
\end{figure}

To address the reliance on global information, the complexity of learning in a hybrid action space in existing methods, and the risk of overloading the network, we propose a priority-driven multi-agent reinforcement learning (MARL) algorithm in which each agent jointly learns control policies and communication priorities. Communication slots are then allocated to agents with the highest priorities. The general framework is shown in Figure \ref{fig:sys}. Although the use of continuous priorities has been explored in linear model-based settings~\citep{mastrangelo2019predictive, mager2022scaling}, our algorithm jointly learns control policies and communication priorities from data. This algorithm offers three significant advantages. First, by focusing on continuous communication priorities, we simplify the hybrid action space in existing methods to a purely continuous one. Second, the priority-based algorithm is well-suited to bandwidth-limited settings, since it allocates communication slots to agents with the highest priorities, thereby avoiding network overload. Most importantly, our plug-and-play framework can be combined with any decentralized MARL algorithm, requiring only a few adjustments.

\section{Problem Setting}
This section introduces the multi-agent system we study and formulates the problem as a Decentralized Partially Observable Markov Decision Process (DEC-POMDP) for learning both control and communication policies.
\subsection{System Overview}
Figure~\ref{fig:sys} illustrates the overall framework. We consider a decentralized multi-agent system with $N$ agents, where the dynamics of agent $i$ are
\begin{equation}
    o_i^{t+1} = f_i(o_i^t, \hat{o}^t, a_i^t, \eta_i^t),
\end{equation}
with time step $t \in \mathbb{N}$, local observation $o_i^t \in \mathbb{R}^n$, communicated information $\hat{o}^t$, action $a_i^t \in \mathbb{R}^m$, and process noise $\eta_i^t \in \mathbb{R}^n$. Each agent learns a policy $\pi(a_i \mid o_i,\hat{o})$ based on its local observation and information received to make optimal decisions. In our framework, the action consists of two components: the control action $a_{i,\mathrm{c}}$ and the communication priority $a_{i,\mathrm{p}}$. The communication priorities of all agents determine the communication policy.

To illustrate how communication priorities determine communication policies under limited bandwidth, we consider the case in which only $L$ ($L < N$) agents can transmit information simultaneously. To manage this, agents need to exchange their communication priorities through the network to determine which $L$ agents hold the highest priorities. The key idea for exchanging priorities and determining the communication policy is to exploit the size advantage of priorities. Priorities can be cast as unsigned integer values, while the observation is typically a vector with multiple floating-point numbers. Thus, adding the priorities to the messages containing the observations that are exchanged anyway adds limited overhead. A potential real-world implementation is presented by~\citet{mager2022scaling}. Their many-to-all primitive broadcasts a priority aggregate through the network so that, at the end of each round, the aggregate is available to all agents. Thus, agents can locally determine whether their priority is among the $L$ highest priorities. The final communication policy, therefore, allocates bandwidth to the \(L\) agents with the highest priorities, as shown in the flow from the ``Priorities Exchange'' block to the wireless network and back to the agents. The ``Observations Exchange'' block in Figure~\ref{fig:sys} indicates that agents with the highest priorities transmit their observations \(\hat{o}\), enabling others to update their policies with additional information. Consequently, the communicated information consists of two components: concatenation of $L$ agents' communicated observations $\hat{o}^t$ and communication priorities. To ensure a consistent policy input, communicated observations are arranged in a fixed agent-ID order, with zero vectors padded for non-communicating agents and for the receiving agent itself.

\subsection{Decentralized Partially Observable Markov Process}
We model the multi-agent system as a DEC-POMDP,
$\langle S,\mathcal{A},P,R,O,N,\gamma\rangle$, where $S\subseteq\mathbb{R}^s$ is the state space, $\mathcal{A}=\prod_{i=1}^N \mathcal{A}_i \subseteq \mathbb{R}^{m \times N}$ is the joint action space, $P:\,S \times \mathcal{A} \times S'\rightarrow [0, 1]$ is the transition probability, $R:\,S\times\mathcal{A} \to \mathbb{R}$ is the shared reward function, $O=\prod_{i=1}^N O_i\subseteq \mathbb{R}^{n \times N}$ is the joint local observation space, $N$ is the number of agents, and $\gamma\in(0,1)$ is the discount factor. Each agent's action consists of a control action $a_{i,\mathrm{c}}$ and a communication priority $a_{i,\mathrm{p}}$, so that $\mathcal{A}_i=\mathcal{A}_{i,\mathrm{c}}\times\mathcal{A}_{i,\mathrm{p}}$. Since the system is decentralized, agent $i$ relies only on its local observation $o_i$ and communicated information $\hat{o}$ rather than the global state $s$, and learns a factorized policy
\begin{equation}
   \pi(\mathbf{a} \mid \mathbf{o}, \hat{o}) 
= \prod_{i=1}^N 
\pi_{\mathrm{c}}(a_{i,\mathrm{c}} \mid o_i, \hat{o})\,
\pi_{\mathrm{p}}(a_{i,\mathrm{p}} \mid o_i, \hat{o}),
\end{equation}
 where bold symbols indicate joint vectors of all agents. Only the $L$ agents with the highest communication priorities are allocated communication slots, and these selected agents broadcast their information to all others. Unlike standard RL, where the reward function $R(s, \mathbf{a})$ typically quantifies only the control reward $r_{i,\mathrm{c}}$, we additionally incorporate a communication penalty $r_{i,\mathrm{p}}$ to discourage agents from always assigning high communication priorities. The overall discounted return for agent $i$ is

\begin{equation}
\label{eq:reward}
    \begin{aligned}
 G_i &= \mathbb{E}\left[\sum_{k=0}^\infty \gamma^{k} (r_{i,\mathrm{c}}^{t+k} + r_{i,\mathrm{p}}^{t+k})\right] \\ 
 &= \mathbb{E}\left[\sum_{k=0}^\infty \gamma^{k} (r_{i,\mathrm{c}}^{t+k} - \xi
 a_{i,\mathrm{p}}^{t+k})\right],
\end{aligned}
\end{equation}
where \(a_{i,\mathrm{p}}\) is the communication priority, and $\xi > 0$ is a weighting parameter. For the control reward, we adopt the shared reward setting, a common setting in decentralized MARL algorithms \citep{DPO}. In this setting, agents’ rewards are averaged to yield a single mean reward, and the concrete form of the control reward depends on the considered system. We aim to jointly optimize the control policy and communication priority to maximize the expected
discounted return in~\eqref{eq:reward}.

\section{Methodology}
To optimize performance and facilitate implementation, we choose the Independent Proximal Policy Optimization (IPPO) algorithm \citep{dewitt2020independent} as the base MARL algorithm, integrating it with our framework to form Control-Priority IPPO, while noting that our framework is compatible with other MARL algorithms. In our implementation, we extend the original action space to include a continuous communication priority \(a_{i,\mathrm{p}} \in (0, 1)\), alongside designing a communication priority penalty $r_{i,\mathrm{p}}$. A higher value of \(a_{i,\mathrm{p}}\) indicates a greater urgency and importance of an agent's communication at the expense of a higher communication penalty. This setup allows our framework to seamlessly integrate into IPPO, enabling the joint learning of control policies and communication priorities without additional neural networks. Below, we provide a brief overview of Control-Priority IPPO. It maintains the robustness and simplicity of PPO while allowing agents to learn independently from local observations and communicated information, making it particularly suitable for decentralized settings. For each agent, the actor network, parameterized by $\theta$, learns the control policy and communication priority, while the critic network, parameterized by $\phi$, learns the value function.

The objective for updating the policy of agent $i$ is
\begin{equation}
\label{eq:af}
    \begin{aligned}
       \mathcal{L}_i(\theta) =& \mathbb{E}_t \biggl[ \min \biggl(  \frac{\pi_{\theta}(a_i^t | \Tilde{o}^t_i)}{\pi_{\theta_{old}}(a_i^t | \Tilde{o}^t_i)} A^t_i, \\
& \text{clip}\left( \frac{\pi_{\theta}(a_i^t | \Tilde{o}^t_i)}{\pi_{\theta_{old}}(a_i^t | \Tilde{o}^t_i)}, 1 - \epsilon, 1 + \epsilon \right) A^t_i \biggr) \biggr],  
    \end{aligned}
\end{equation}
where $\mathbb{E}_t[\cdot]$ denotes the empirical expectation over the sampled trajectories, $\epsilon$ is the clipping ratio (set to 0.2), $\Tilde{o}^t_i$ is the concatenation of local observation $o^t_i$ and information $\hat{o}^t$ from agents who are allowed to communicate, and $A_i^t$ is estimated by 
Generalized Advantage Estimation (GAE) \citep{DBLP:journals/corr/SchulmanMLJA15}. 
The clipped surrogate follows \citet{schulman2017proximal}. Specifically,
\begin{equation}
\label{eq:gae}
    A^t_i = \sum_{k=0}^{T} (\gamma\lambda)^k \delta^{t+k}_i, 
\end{equation}
where $\lambda$ is the discount factor for GAE, $T$ is the time steps for data collection, $\delta^t_i = r^t_i + \gamma V_\phi(\Tilde{o}^{t+1}_i) - V_\phi(\Tilde{o}^t_i)$ and $V_\phi$ is the value function approximated by the critic network. The objective for updating the value function of agent $i$ is
\begin{equation}
\label{eq:vf}
    \begin{aligned}
         &\mathcal{L}_i(\phi) = \mathbb{E}_t\biggl[ \min \biggl( \left(V_\phi(\Tilde{o}^t_i) - \hat{V}_i^t \right)^2, \\
    &\left(V_{\phi_{old}}(\Tilde{o}^t_i) + \text{clip}\left(V_\phi(\Tilde{o}^t_i) - V_{\phi_{old}}(\Tilde{o}^t_i), -\epsilon, \epsilon\right) - \hat{V}_i^t \right)^2 \biggl) \biggl],
    \end{aligned}
\end{equation}
where \(\phi_{old}\) are old parameters before the update, and \(\hat{V}_i^t = A_i^t + V_\phi(\Tilde{o}^t_i)\) \citep{dewitt2020independent}. The learning procedure of our method is summarized in Algorithm~\ref{alg:PIPPO}. For each agent, the rollout buffer $D_i$ stores the samples collected. After each rollout, GAE is computed from $D_i$, and mini-batches sampled from $D_i$ are used to update the actor and critic networks according to Equations~\eqref{eq:af} and~\eqref{eq:vf}.
\begin{algorithm2e}
\caption{Control-Priority Independent Proximal Policy Optimization}
\label{alg:PIPPO}
\SetAlgoLined
\SetAlgoNoEnd
Initialize $\theta_i,\phi_i$ for all agents $i=1,\dots,N$\;
\While{step $\leq$ step$_{\max}$}{
    $D_i \leftarrow \emptyset$ for all agents\;
    \For{$t = 1$ \KwTo $T$}{
        \For{$i = 1$ \KwTo $N$}{
            $a^t_{i,\mathrm{c}}, a^t_{i,\mathrm{p}} \sim \pi_{\theta}(o_i^t,\hat{o}^t)$,\quad
            $v_i^t \leftarrow V_{\phi}(o_i^t,\hat{o}^t)$\;
            Execute $a^t_{i,\mathrm{c}}$, observe $r^t_{i,\mathrm{c}}, {o}_i^{t+1}$\;
            Store $(a^t_{i,\mathrm{c}}, a^t_{i,\mathrm{p}}, v_i^t, r^t_{i,\mathrm{c}}, \Tilde{o}_i^{t}, o_i^{t+1})$ in $D_i$\;
        }
         The top-$L$ agents transmit $\hat{o}^{t+1}$ based on $\pi^t_{\mathrm{p}}$ and all agents send priorities\;
        Observe $\mathbf{r}^t_{\mathrm{p}}$ and store $(r^t_{i,\mathrm{p}},\hat{o}^{t+1})$ in each $D_i$\;
    }
    Compute and store $\mathbf{A}$ via Equation~\eqref{eq:gae}\;
    \For{$k = 1$ \KwTo $K$}{
        \For{$i = 1$ \KwTo $N$}{
            Sample $B_i$ from $D_i$ and update $\theta_i,\phi_i$ using
            Equations~\eqref{eq:af} and~\eqref{eq:vf}\;
        }
    }
}
\end{algorithm2e}

\section{Experiments}
In this section, we present experimental results\footnote{Videos and code of our experimental results are available at \url{https://etcpriority.github.io/}.} in the following simulation environments: Multiwalker \citep{terry2021pettingzoo}, Coverage Control, and Formation Control \citep{agarwal2019learning}. These environments span both low- and high-dimensional scenarios and involve varying numbers of agents, allowing us to comprehensively evaluate the effectiveness of our algorithm. We use IPPO with round-robin communication as the baseline. In this setting, agents communicate in turns, learn only control policies, and therefore incur no communication penalty. Following \citet{mager2022scaling}, we allocate $L_1$ communication slots to round-robin communication and $L_2$ slots to priority communication, with $L_1 \geq L_2$. The smaller number of slots for our algorithm accounts for the additional overhead of transmitting compact priority values. We train 10 million steps for all environment settings. Solid lines in training figures are mean episode rewards over three random seeds, and the shaded area is the standard deviation. 
\subsection{Environments Description}
\subsubsection{Coverage Control}
\label{sec:Coverage}
In Coverage Control, there are $N$ agents and $M$ landmarks, as shown in Figure~\ref{fig:converge}. The goal is for agents to position themselves such that each landmark is within a certain distance of at least one agent. The control action is two-dimensional, corresponding to horizontal and vertical motion. The shared reward is the negative average distance from each landmark to its nearest agent~\citep{agarwal2019learning}. In the native implementation, the local observation of an agent includes its position, velocity, the position of landmarks relative to the agent, and the positions of other agents relative to the agent. To make communication necessary, we remove the information of other agents, which we obtain through communication. Thus, the local observation of each agent consists of its position, velocity, and the relative position of landmarks. The communicated information is the other agents' positions and velocities. 

\subsubsection{Formation Control}
\label{sec:formation}
\begin{figure}[tbp]
\centering
\begin{subfigure}{0.18\textwidth}
   \centering
    \includegraphics[width=0.75\linewidth]{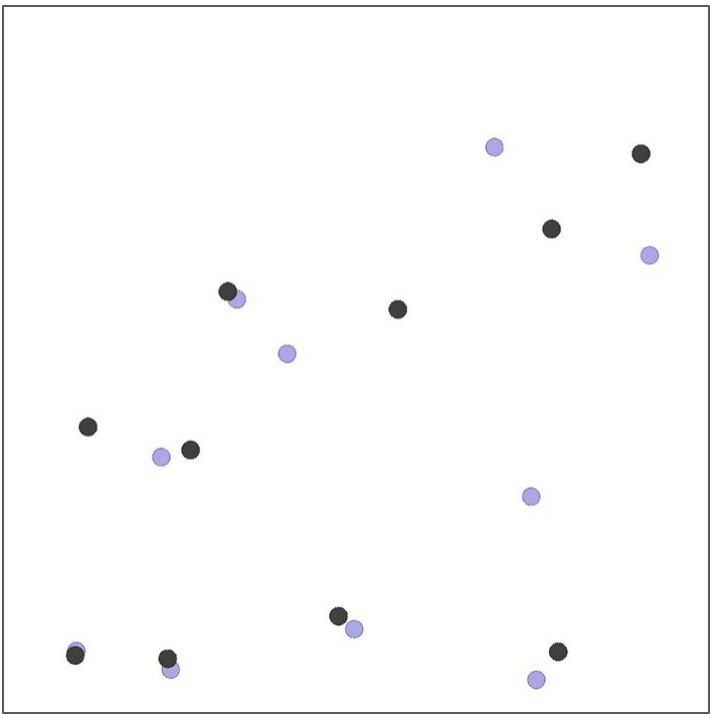}
    \caption{Coverage Control}
    \label{fig:converge}
\end{subfigure}
\begin{subfigure}{0.18\textwidth}
    \centering
    \includegraphics[width=0.75\linewidth]{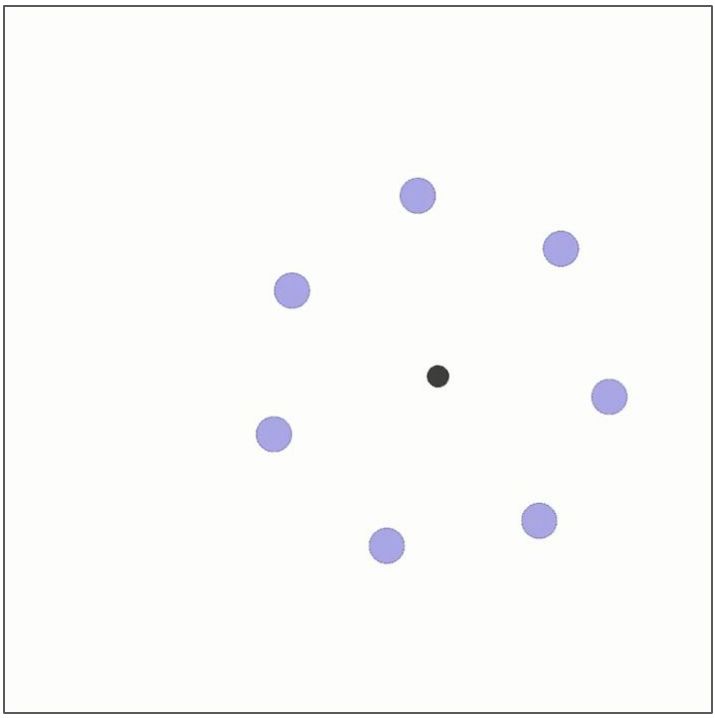}
    \caption{Formation Control}
    \label{fig:formation}
\end{subfigure}
     \caption{(a) Coverage Control: Agents (purple) deploy themselves so that every agent reaches a distinct landmark (black). (b) Formation Control: Agents (purple) position themselves into a regular polygonal formation, with the landmark (black) at its center.} 
\end{figure}

There are $N$ agents and one landmark in this environment, as shown in Figure \ref{fig:formation}. The objective is for agents to form an $N$-sided regular polygon with the landmark at its center. The orientation of the polygon is not specified, nor are agents assigned to specific points on the polygon. Instead, the agents communicate with each other to distribute themselves uniformly \citep{agarwal2019learning}. The control action space in this environment is the same as in Coverage Control. The reward function for this environment consists of two components: a penalty for deviation from the target distance of 0.5 units to the landmark and a penalty for deviation from the ideal angular spacing of $\frac{2\pi}{N}$ between agents. We make the same modifications to local observations in this environment as in Coverage Control. 

\subsubsection{Multiwalker}
\label{sec:multiwalker}
\begin{figure}[tbp]
\centering
    \includegraphics[width=0.45\linewidth]{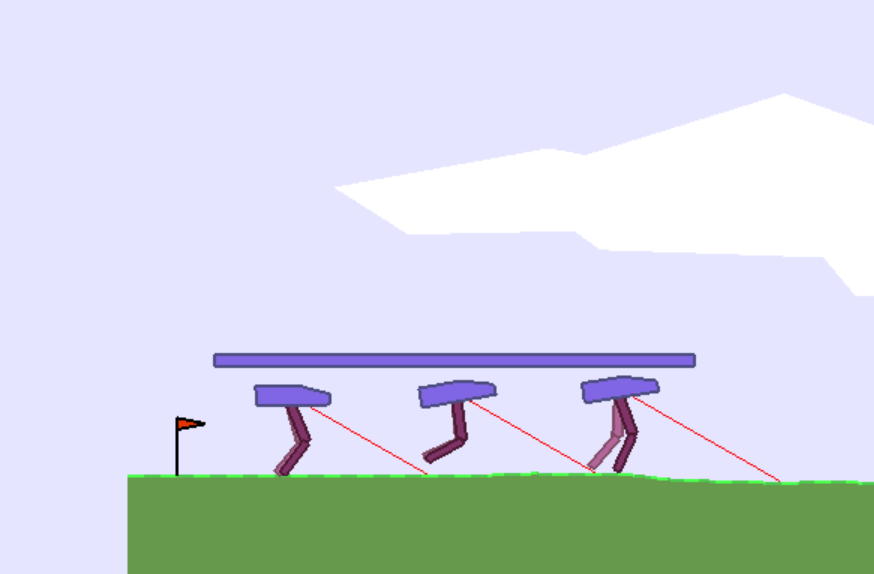}
    \caption{Multiwalker environment: Walkers try to transport the package to
	the destination without falling down.}
    \label{fig:multiwalker}
\end{figure}

We select this environment to evaluate our algorithm's performance in a high-dimensional setting. In this environment, bipedal robots (walkers) aim to transport a package from the left side to the right side of the terrain; see Figure \ref{fig:multiwalker}. Each walker applies force to two joints in each of its two legs, resulting in a continuous action space represented by a 4-element vector. Walkers get a reward based on the distance the package moves forward each time step. The simulation ends if the package falls off the walkers, goes off the left edge of the terrain, or if any walker falls. In these cases, each walker receives a penalty of -100.

In the native implementation of Multiwalker, the observation includes the agent's joint status, LIDAR information about the environment in front of them, state information about their neighbors in the form of relative positions, and their relative position and angle towards the package. To make communication necessary, we follow the modifications by \citet{pmlr-v211-kesper23a} and remove all LIDAR information and information of neighbors. Each agent can only receive information about its joint status and the package. When agent $i$ communicates, it sends its absolute position and velocity to other agents, from which receiving agents can calculate relative positions and velocities.

\subsubsection{Network Model and Message Size}
For both our algorithm and IPPO-Round Robin, we use the same network disturbance settings in all environments: message loss is uniformly distributed with probability 0.2, and message delay is one time step. If an agent's message is dropped, it is not delivered to any other agent. When priority messages are lost, our algorithm defaults to round-robin communication for that time step. In all environments, each communicated observation contains the agent's positions and velocities along the $x$- and $y$-axes, represented by four \SI{32}{\bit} floating-point numbers. So, the size of one message is \SI{16}{\byte}. By contrast, communication priorities can be encoded as small integers and require only about \SI{3}{\bit} in our settings.



\subsection{Results}

\subsubsection{N=3, $L_1$=1, $L_2$=1}
We set the number of agents to three and allow one agent to communicate per time step. Under this setting, we test our algorithm in Coverage Control and Multiwalker.

\begin{figure*}[tbp]
\centering

\begin{subfigure}[t]{0.37\textwidth}
    \centering
    \includegraphics[width=\linewidth,trim=2 2 2 2,clip]{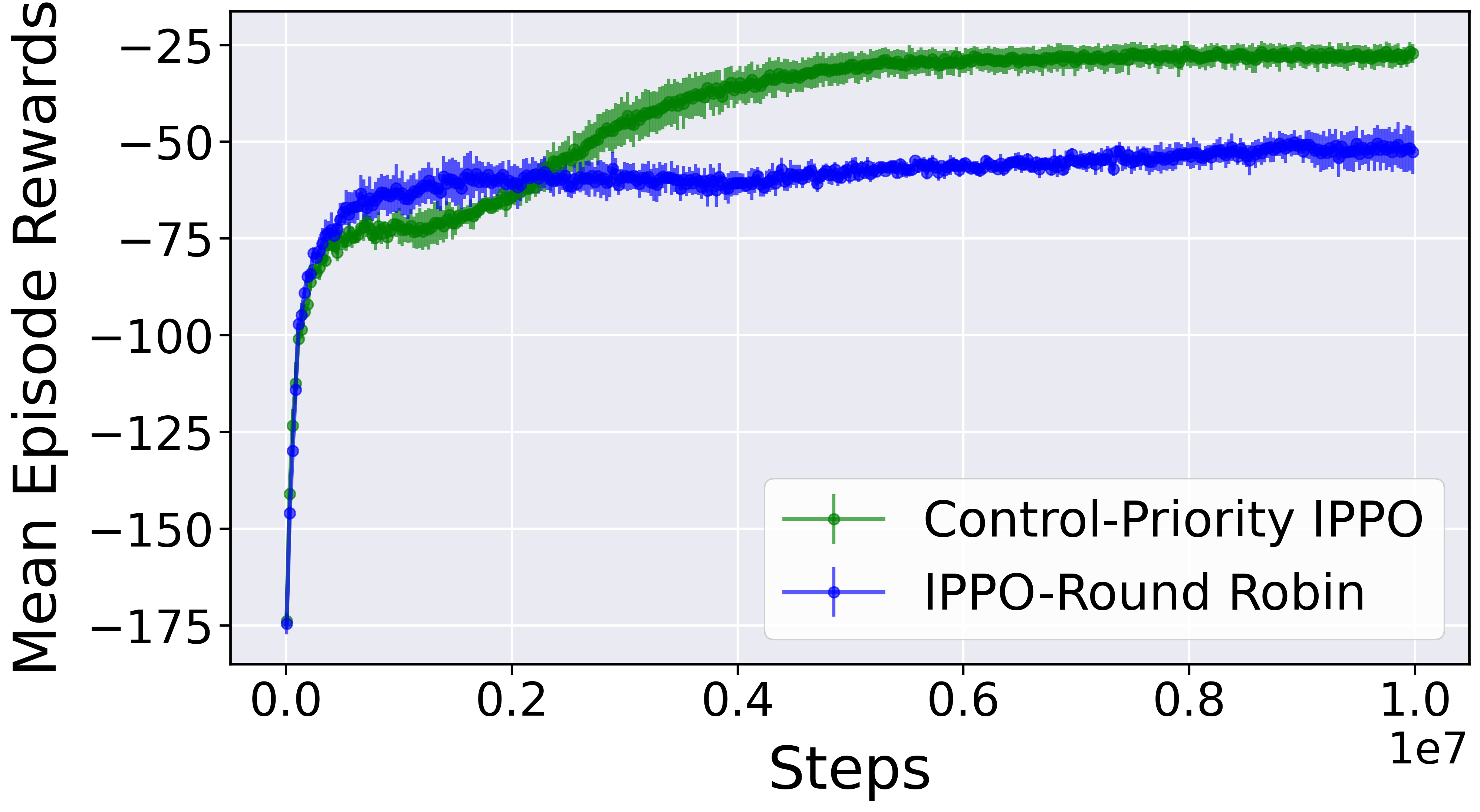}
    \caption{Coverage Control with three agents.}
    \label{fig:re1}
\end{subfigure}
\hspace{0.1\textwidth}
\begin{subfigure}[t]{0.37\textwidth}
    \centering
    \includegraphics[width=\linewidth,trim=3 3 3 3,clip]{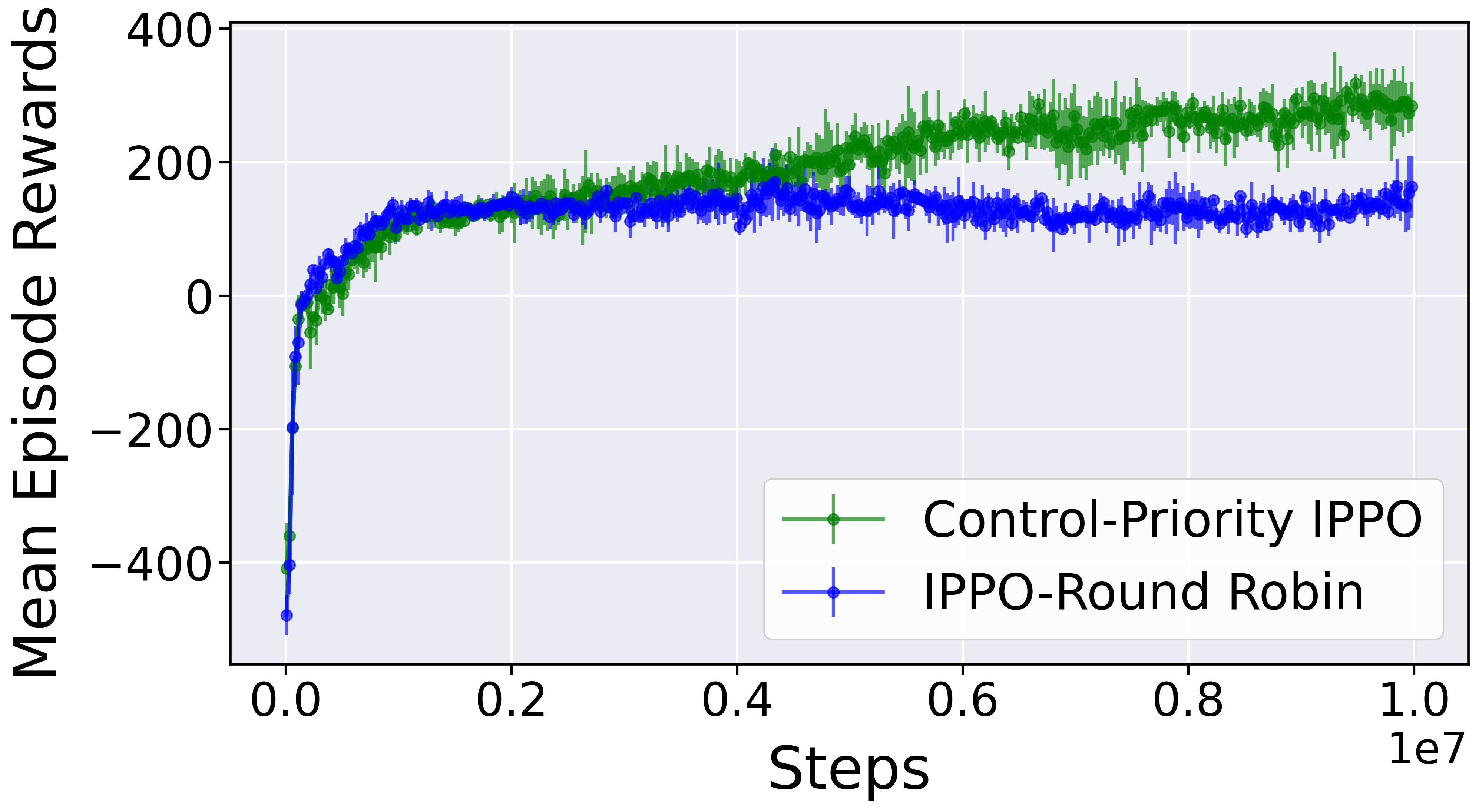}
    \caption{Multiwalker with three agents.}
    \label{fig:re2}
\end{subfigure}

\vspace{0.3em}

\begin{subfigure}[t]{0.37\textwidth}
    \centering
    \includegraphics[width=\linewidth,trim=3 3 3 3,clip]{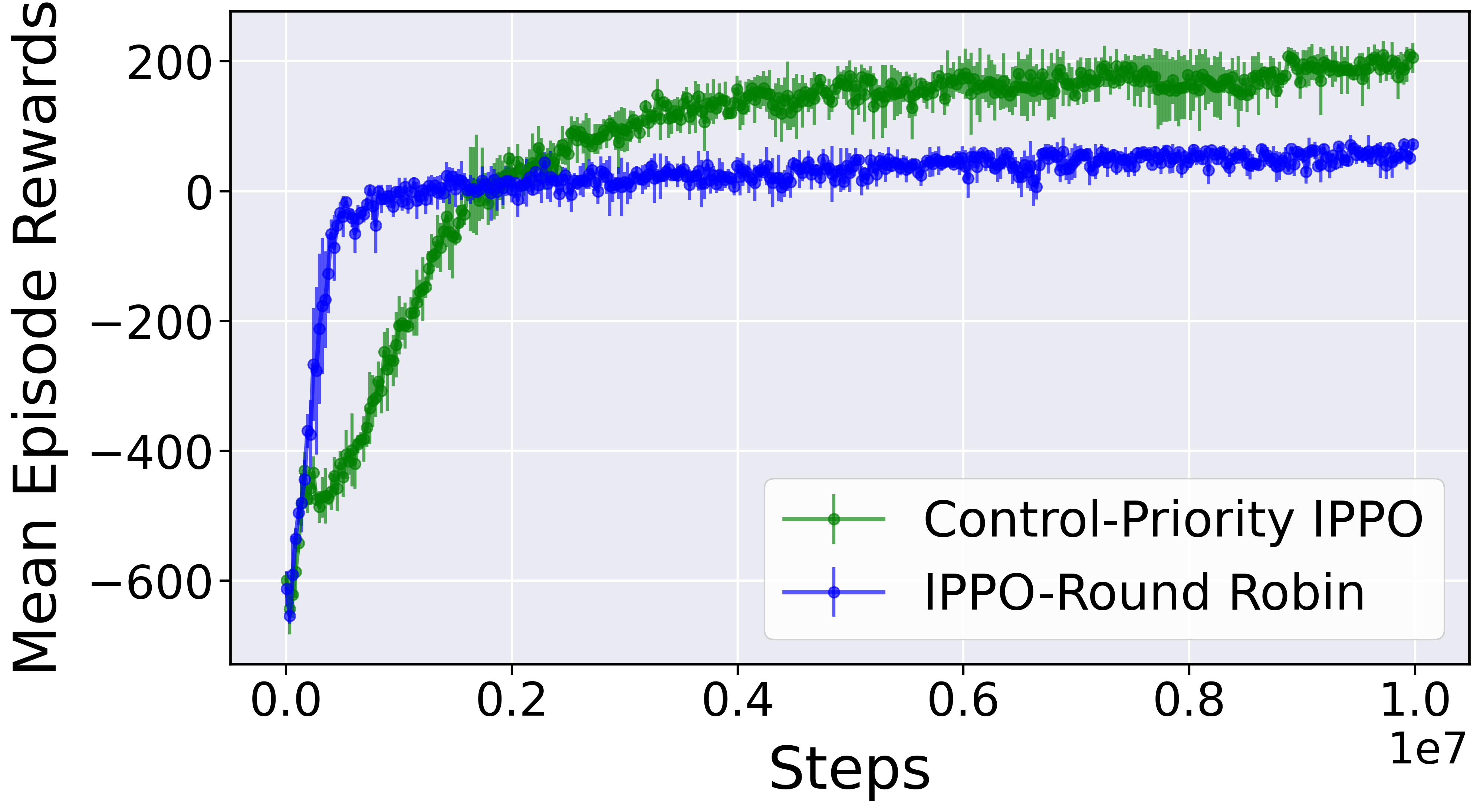}
    \caption{Multiwalker with six agents.}
    \label{fig:re3}
\end{subfigure}
\hspace{0.1\textwidth}
\begin{subfigure}[t]{0.37\textwidth}
    \centering
    \includegraphics[width=\linewidth,trim=3 3 3 3,clip]{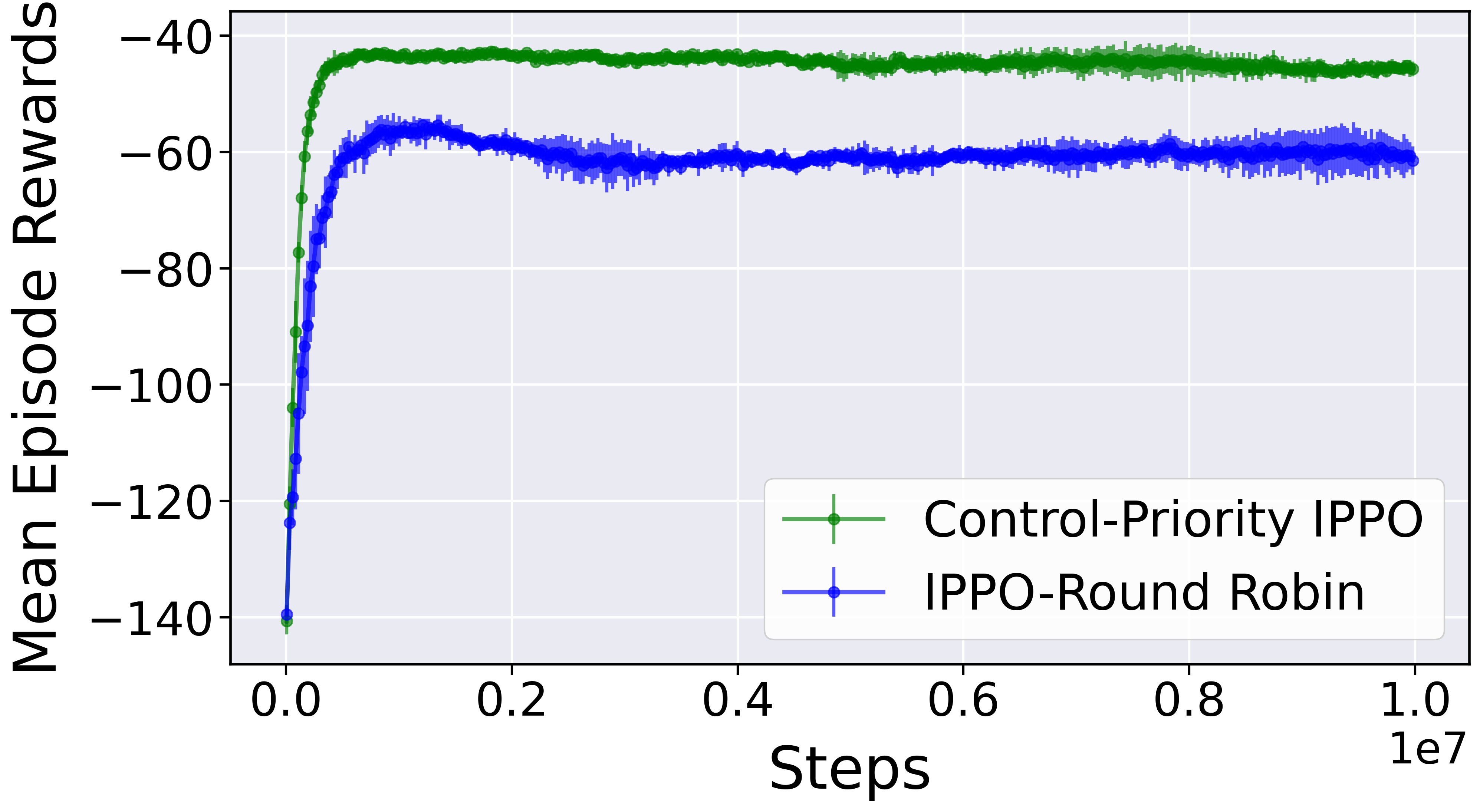}
    \caption{Formation Control with eight agents.}
    \label{fig:re4}
\end{subfigure}

\caption{Results for four tasks and agent configurations. Our algorithm (green) outperforms the baseline (blue) in all evaluated settings. Episode rewards of our algorithm include the communication penalty.}
\label{fig:images}
\end{figure*}

For Coverage Control, as shown in Figure \ref{fig:re1}, IPPO with round-robin communication converges faster because it does not need to learn communication policies. However, naive round-robin communication limits the ability to learn better control policies. Our algorithm requires more training steps to balance communication penalties and communication needs. Still, it eventually achieves higher average rewards than IPPO with round-robin communication. Under bandwidth constraints, our algorithm allows the agent with the highest priority to communicate, making the transmitted information more useful for other agents to learn control policies than in round-robin communication. 

For Multiwalker, Figure~\ref{fig:re2} shows a clear advantage of our method over IPPO with round-robin communication after the initial learning phase. This suggests that priority-based communication can improve control performance in nonlinear and high-dimensional environments.

\begin{figure*}[tbp]
\centering
\begin{overpic}[width=0.62\textwidth]{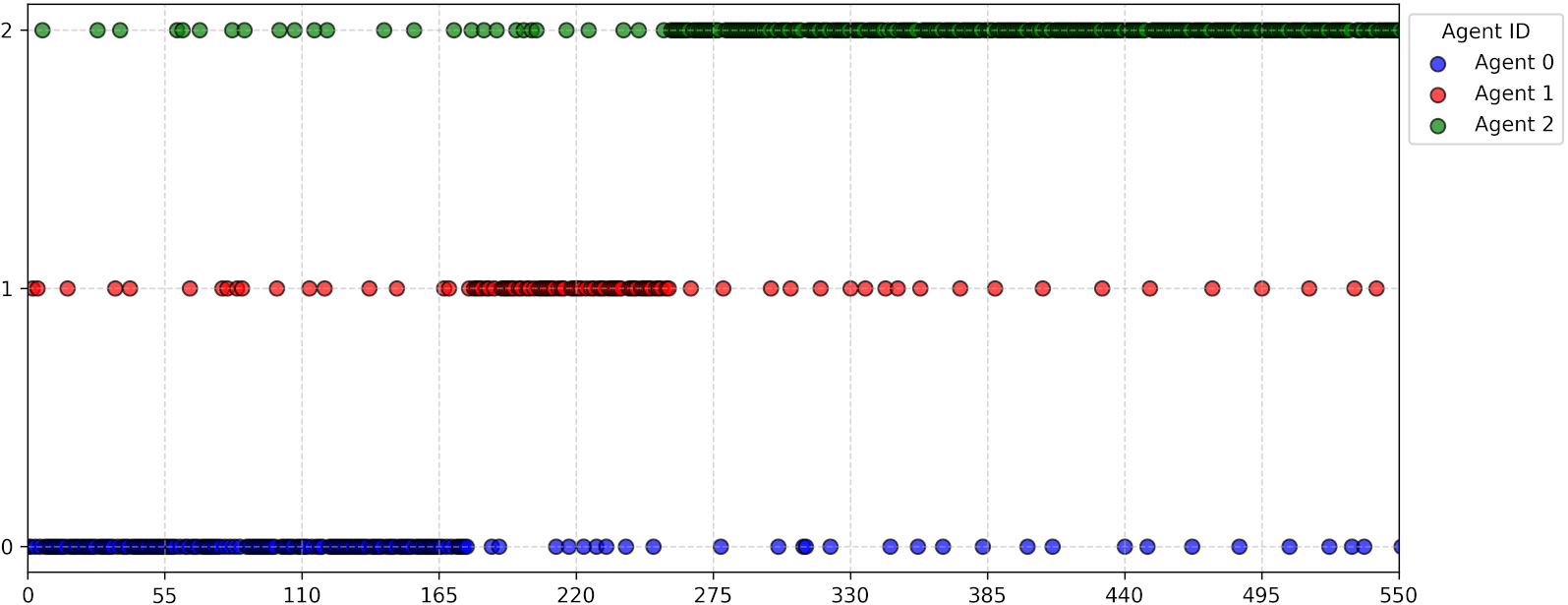}
    \put(43,-2.5){\scriptsize Steps}
    \put(-3.5,7.5){\rotatebox{90}{\scriptsize Highest-Priority Agent}}
\end{overpic}
\vspace{0.2em}

\caption{Communication priority dynamics in the evaluation of the three-agent Multiwalker scenario.}
\label{fig:walker_p}
\end{figure*}

\subsubsection{N=6, $L_1$=2, $L_2$=1}
To further evaluate both the effectiveness and scalability of our algorithm under restricted bandwidth, we increase the number of agents to six and allocate more communication slots to round-robin communication, i.e., two slots for round-robin and one for our algorithm. Under this setting, we test our algorithm in Multiwalker.

From Figure \ref{fig:re3}, we can see that when the number of agents increases to six, both methods recover quickly from initial large negative rewards, representing early trial and error where agents learn basic coordination without causing the package to fall. The episode rewards of IPPO with round-robin communication converge to approximately 70. Although our algorithm requires longer training time to converge, agents with priority communication achieve significantly higher performance, with episode rewards converging to over 200, and can move the package farther than the baseline. Comparing the two Multiwalker settings with the same communication slot ($L_2=1$), the converged episode rewards of our method decrease from approximately 290 in the three-agent scenario to around 200 in the six-agent scenario, indicating that coordination becomes more challenging as the number of agents increases relative to the available bandwidth.

\subsubsection{N=8, $L_1$=2, $L_2$=1}
To show that our algorithm still has an advantage over round-robin communication with more agents and restricted bandwidth, we increase the number of agents to eight and allocate two communication slots to round-robin communication and one to our algorithm.

We choose a low-dimensional environment, Formation Control, to test our algorithm to ensure that IPPO can learn good control policies with eight agents. From Figure \ref{fig:re4}, we can see that over 10 million steps, our algorithm consistently outperforms IPPO with round-robin communication. This suggests that our algorithm is even more effective with more agents and restricted bandwidth, leveraging its single communication slot to optimize performance in a more crowded environment. Despite having more communication slots, IPPO with round-robin communication does not perform better, suggesting that merely increasing communication frequency without strategic prioritization is less effective in this environment.
\subsubsection{Communication Priority Evaluation in Multiwalker}

To show that our algorithm learns meaningful communication priorities, we present the communication-priority dynamics of the three-agent Multiwalker scenario in Figure~\ref{fig:walker_p}. The highest-priority agent changes over time according to the current task phase: agent \SI{0} is prioritized when initiating the lift and handover, agent \SI{1} is prioritized during the intermediate transfer, and agent \SI{2} is prioritized in the final transport stage. This indicates that the learned priorities reflect the agents' time-varying roles in coordination.

\subsubsection{Discussion of Scalability and Bandwidth Efficiency}
Compared to MAPPOC~\citep{pmlr-v211-kesper23a}, our algorithm offers advantages in both bandwidth efficiency and scalability. In the three-agent Multiwalker setting, MAPPOC reports \SI{55}{\percent} bandwidth savings~\citep{pmlr-v211-kesper23a}, whereas our method uses only one communication slot per time step, achieving about \SI{66}{\percent} savings with higher rewards. In addition, as a CTDE-based method, MAPPOC requires global information during training, which causes the state space to grow with the number of agents and makes value-function learning more difficult~\citep{Qin2021LearningSM}. This issue becomes more pronounced when the number of agents increases and bandwidth is limited. In addition, our experiments account for message delays and losses, whereas many existing methods assume ideal communication~\citep{pmlr-v211-kesper23a, Baumann2018DeepRL}.

\section{Conclusion}
In this paper, we present a priority-driven algorithm for the joint learning of control and communication policies in a decentralized manner. Our approach has three advantages: (i) it simplifies the complexity of the hybrid action space, (ii) its priority-based communication mechanism can be easily adjusted to different bandwidth constraints, thereby improving flexibility, and (iii) it maintains a plug-and-play design, requiring minimal modifications to existing MARL algorithms.

Our algorithm achieves strong results while reducing bandwidth usage and maintaining high control performance under delay and message loss. Its scalability is improved by relying only on local observations and communicated information and by allocating bandwidth according to learned priorities, but the overall scalability remains limited by the underlying MARL algorithm. Extending this framework to larger numbers of agents and other decentralized MARL methods is an important direction for future work.

\begin{ack}
    We acknowledge the computational resources provided by the Aalto Science-IT project.
\end{ack}

\bibliography{ifacconf}             

@inproceedings{Heemels2012AnIT,
  title={An introduction to event-triggered and self-triggered control},
  author={Heemels, W.P.M.H. and Johansson, Karl Henrik and Tabuada, Paulo},
  booktitle={IEEE Conference on Decision and Control},
  year={2012},
}

@book{Miskowicz2015,
  editor       = {Miskowicz, Marek},
  title        = {Event-Based Control and Signal Processing},
  publisher    = {CRC Press},
  year         = {2015},
  edition      = {1}
}

@article{FUNK2021100144,
title = {Learning event-triggered control from data through joint optimization},
journal = {IFAC Journal of Systems and Control},
year = {2021},
author = {Niklas Funk and Dominik Baumann and Vincent Berenz and Sebastian Trimpe},

}

@INPROCEEDINGS{5990925,
  author={Ramesh, Chithrupa and Sandberg, Henrik and Bao, Lei and Johansson, Karl Henrik},
  booktitle={Proceedings of the American Control Conference}, 
  title={On the dual effect in state-based scheduling of networked control systems}, 
  year={2011},}

@InProceedings{pmlr-v211-kesper23a,
  title = 	 {Toward Multi-Agent Reinforcement Learning for Distributed Event-Triggered Control},
  author =       {Kesper, Lukas and Trimpe, Sebastian and Baumann, Dominik},
  booktitle = 	 {Learning for Dynamics and Control Conference},
  year = 	 {2023},
}

@article{SHIBATA2023104307,
title = {Deep reinforcement learning of event-triggered communication and consensus-based control for distributed cooperative transport},
journal = {Robotics and Autonomous Systems},
year = {2023},
author = {Kazuki Shibata and Tomohiko Jimbo and Takamitsu Matsubara},
}

@article{DPO,
title={A Fully Decentralized Surrogate for Multi-Agent Policy Optimization},
author={Su, Kefan and Lu, Zongqing},
journal={Transactions on Machine Learning Research},
year={2024}
}

@article{schulman2017proximal,
  title={Proximal Policy Optimization Algorithms},
  author={John Schulman and Filip Wolski and Prafulla Dhariwal and Alec Radford and Oleg Klimov},
  journal={ArXiv},
  year={2017},
  volume={abs/1707.06347}
}

@inproceedings{DBLP:journals/corr/SchulmanMLJA15,
  author       = {John Schulman and
                  Philipp Moritz and
                  Sergey Levine and
                  Michael I. Jordan and
                  Pieter Abbeel},
  title        = {High-Dimensional Continuous Control Using Generalized Advantage Estimation},
  booktitle    = {International Conference on Learning Representations},
  year         = {2016}
}

@inproceedings{terry2021pettingzoo,
 author = {Terry, J and Black, Benjamin and Grammel, Nathaniel and Jayakumar, Mario  and Hari, Ananth  and Sullivan, Ryan and Santos, Luis S and Dieffendahl, Clemens and Horsch, Caroline and Perez-Vicente, Rodrigo and Williams, Niall  and Lokesh, Yashas  and Ravi , Praveen },
 booktitle = {Advances in Neural Information Processing Systems},
 title = {PettingZoo: Gym for Multi-Agent Reinforcement Learning},
 year = {2021}
}

@article{dewitt2020independent,
  title={Is Independent Learning All You Need in the StarCraft Multi-Agent Challenge?},
  author={C. S. D. Witt and Tarun Gupta and Denys Makoviichuk and Viktor Makoviychuk and Philip H. S. Torr and Mingfei Sun and Shimon Whiteson},
  journal={ArXiv},
  year={2020},
  volume={abs/2011.09533}
}

@inproceedings{agarwal2019learning,
author = {Agarwal, Akshat and Kumar, Sumit and Sycara, Katia and Lewis, Michael},
title = {Learning Transferable Cooperative Behavior in Multi-Agent Teams},
year = {2020},
booktitle = {International Conference on Autonomous Agents and MultiAgent Systems},

}

@ARTICLE{9830835,
  author={Sedghi, Leila and Ijaz, Zohaib and Noor-A-Rahim, Md. and Witheephanich, Kritchai and Pesch, Dirk},
  journal={IEEE Access}, 
  title={Machine Learning in Event-Triggered Control: Recent Advances and Open Issues}, 
  year={2022},
  }

@inproceedings{
Qin2021LearningSM,
title={Learning Safe Multi-agent Control with Decentralized Neural Barrier Certificates},
author={Zengyi Qin and Kaiqing Zhang and Yuxiao Chen and Jingkai Chen and Chuchu Fan},
booktitle={International Conference on Learning Representations},
year={2021},
}

@ARTICLE{8663430,
  author={Yang, Xiong and He, Haibo},
  journal={IEEE Transactions on Systems, Man, and Cybernetics: Systems}, 
  title={Adaptive Critic Learning and Experience Replay for Decentralized Event-Triggered Control of Nonlinear Interconnected Systems}, 
  year={2020},
  
  }

@ARTICLE{8713542,
  author={Tan, Luy Nguyen},
  journal={IEEE Transactions on Systems, Man, and Cybernetics: Systems}, 
  title={Event-Triggered Distributed {H}$\infty$ Constrained Control of Physically Interconnected Large-Scale Partially Unknown Strict-Feedback Systems}, 
  year={2021},
  
  }

@article{mastrangelo2019predictive,
  title={Predictive triggering for distributed control of resource constrained multi-agent systems},
  author={Mastrangelo, Jos{\'e} Mario and Baumann, Dominik and Trimpe, Sebastian},
  journal={IFAC-PapersOnLine},
  
  year={2019},
}

@article{mager2022scaling,
  title={Scaling beyond bandwidth limitations: Wireless control with stability guarantees under overload},
  author={Mager, Fabian and Baumann, Dominik and Herrmann, Carsten and Trimpe, Sebastian and Zimmerling, Marco},
  journal={ACM Transactions on Cyber-Physical Systems},
  
  year={2022},
}

@ARTICLE{8386658,
  author={Demirel, Burak and Ramaswamy, Arunselvan and Quevedo, Daniel E. and Karl, Holger},
  journal={IEEE Control Systems Letters}, 
  title={DeepCAS: A Deep Reinforcement Learning Algorithm for Control-Aware Scheduling}, 
  year={2018},
  }

@article{Baumann_2021,
   title={Wireless Control for Smart Manufacturing: Recent Approaches and Open Challenges},
   
   journal={Proceedings of the IEEE},
   
   author={Baumann, Dominik and Mager, Fabian and Wetzker, Ulf and Thiele, Lothar and Zimmerling, Marco and Trimpe, Sebastian},
   year={2021},
    }

@article{Baumann2018DeepRL,
  title={Deep Reinforcement Learning for Event-Triggered Control},
  author={Dominik Baumann and Jia-Jie Zhu and Georg Martius and Sebastian Trimpe},
  journal={ IEEE Conference on Decision and Control },
  year={2018},
  
}

@article{Dang2022EventTriggeredMP,
  title={Event-Triggered Model Predictive Control With Deep Reinforcement Learning for Autonomous Driving},
  author={Fengying Dang and Dong Chen and Jun Wei Chen and Zhaojian Li},
  journal={IEEE Transactions on Intelligent Vehicles},
  year={2022},
  
  
}








\end{document}